\documentclass[useAMS,usenatbib,usegraphicx]{aa}  
\usepackage{graphicx}
\usepackage{txfonts}
\usepackage{epsfig}
\usepackage{amsmath} 
\usepackage{rotating}           
\usepackage{color}     
\usepackage{graphicx}
\usepackage{upgreek} 

\def\kms{km ${\rm s}^{-1}$}

\def\Mo{M$_\odot$}

\def\ccm {$\hbox{{\rm cm}}^{-3}$}    
\def\scm  {$\hbox{{\rm cm}}^{-2}$}    
\def \AL {$\alpha $}     
\def \HI {H{\sc \,i}}

\def\MOLH {\hbox{${\rm H}_2$}}  

\def\lapp{\ifmmode\stackrel{<}{_{\sim}}\else$\stackrel{<}{_{\sim}}$\fi}
\def\gapp{\ifmmode\stackrel{>}{_{\sim}}\else$\stackrel{>}{_{\sim}}$\fi}

\begin{document} 
\title{The potential of tracing the star formation history with \HI\ 21-cm in intervening absorption systems} 
 \author{S. J. Curran
          }
  \institute{School of Chemical and Physical Sciences, Victoria University of Wellington, PO Box 600, Wellington 6140, New Zealand\\
  \email{Stephen.Curran@vuw.ac.nz} 
             }

 
\abstract{
Unlike the neutral gas density, which remains largely constant over redshifts of $0\lapp z\lapp 5$,
the star formation density, $\psi_{*}$, exhibits a strong redshift dependence, increasing
from the present day before peaking at a redshift of $z\approx2.5$.
Thus, there is a stark contrast between the star formation rate and the abundance of raw material available to fuel it. However,
using the ratio of the strength of the \HI\ 21-cm absorption to the total neutral gas column density to quantify the
spin temperature, $T_{\rm spin}$, of the gas, it has recently been shown that $1/T_{\rm spin}$ may trace
$\psi_{*}$. This would be expected on the grounds that the cloud of gas must be sufficiently cool to collapse under
its own gravity. This, however, relies on very limited data and so here we explore the potential of applying the above
method to absorbers for which individual column densities are not available (primarily Mg{\sc \,ii} absorption systems). By
using the mean value as a proxy to the column density of the gas at a given redshift, we do, again, find that
$1/T_{\rm spin}$ (degenerate with the absorber--emitter size ratio)  traces $\psi_{*}$. If confirmed by higher
redshift data, this could offer a powerful tool for future surveys for cool gas throughout the Universe with the Square
Kilometre Array.
}
  

   \maketitle
%

\section{Introduction}

\label{intro}
 
Neutral hydrogen (\HI), the reservoir for star formation, is traced in the distant Universe through 
21-cm and Lyman-\AL\ absorption by galaxies intervening the sight-line to more distant
radio and optical/UV continuum sources (e.g. \citealt{wgp05}). The majority of this neutral gas (constituting up to 80\%
of the total in the Universe, \citealt{phw05}) arises in the so-called damped Lyman-$\alpha$ absorption systems (DLAs),
defined to have neutral hydrogen column densities of $N_\text{\HI} \ge2\times10^{20}$ \scm.

Since the Lyman-$\alpha$ transition occurs in the ultra-violet band ($\lambda=1216$ \AA), the majority of DLAs are
detected at redshifts of $z_{\rm abs} \gapp1.7$ (e.g. \citealt{npc+12}), where the transition is shifted into the
optical band.  In addition to space-based observations of the Lyman-$\alpha$ transition (e.g.  \citealt{rtsm17}), the
presence of neutral hydrogen at lower redshifts is evident through 21-cm emission studies (currently limited to
$z\lapp0.4$, \citealt{fgv+16}) and may be inferred from the absorption of Mg{\sc \,ii} (e.g. \citealt{rtn05}), or other
low ionised metal species (e.g. \citealt{dsg+17}) which can be observed from the ground. Other intervening absorption
systems not detected in the optical band have been identified through 21-cm and millimetre band molecular absorption
(\citealt{cry93,lrj+96,cdn99,kb03,cdbw07,amm+16}). In these cases, the redshifts are generally too low (currently limited to
$z_{\rm abs}\leq0.96$, \citealt{cdbw07}) and the background continuum sources too optically faint/reddened to yield a
Lyman-$\alpha$ detection \citep{cmpw03,cwm+06}.

From observations of \HI\ 21-cm emission and Lyman-$\alpha$ absorption, both of which give the total neutral hydrogen
column density, the neutral gas mass density of the Universe has been mapped from the present day to redshifts of
$z\sim5$ (look-back times of 12.5 Gyr). This has a value relative to the critical density of $\Omega_{\rm HI}
\approx0.5\times10^{-3}$ at $z\lapp0.5$ \citep{zvb+05,lcb+07,bra12,dsmb13,rzb+13,hsf+15,npr+16}, rising to $\Omega_{\rm
  HI} \approx1\times10^{-3}$ at $z\sim0.5$, where it remains nearly constant over the observed $0.5 \lapp z \lapp 5$
\citep{rt00,ph04,rtn05,cur09a,pw09,npc+12,cmp+15}. Furthermore, the inflow of neutral gas, from within the galaxy or
from the intergalactic medium, which may feed fuel to the star formation sites \citep{mgh+15}, also exhibits a near
constancy with redshift \citep{sm17}. This unchanging abundance of neutral gas is in stark contrast to the steep evolution of the
star formation density, which exhibits a climb, before peaking at $z\sim3$, followed by a decrease at higher redshift
\citep{hb06,bbg+13,ssb+13,lbz+14,md14,zjd+14}.

Thus, there is a clear discrepancy between the star formation history and the reservoir of star forming material
(e.g. \citealt{lbz+14}).  Recently, however, by normalising the strength of the \HI\ 21-cm absorption (which traces the
cold component of the gas) by the column density (which traces all of the neutral gas), \citet{cur17} found evidence for
a similarity between the fraction of cool gas and the star formation density. Although this is physically motivated, as
star formation requires cold, dense, neutral gas (e.g. \citealt{mgh+15}), the presence of which is evident from large
molecular abundances (e.g. \citealt{cw13}), the sample contains only 74 confirmed DLAs and sub-DLAs which have been
searched in 21-cm absorption.  The connection between the cold gas fraction and the star formation density is based
primarily on the both peaks occuring at a similar redshift with a common factor of $\sim10$ over the $z =0$ value. Given
the difficulties in obtaining a large sample of DLAs which exhibit 21-cm absorption\footnote{Given that the majority
  ($\gapp80\%$, \citealt{cwbc01}) of background sources are "radio-quiet" ($\lapp0.1$ Jy for our purposes), the chances
  of finding a sufficiently strong source, where the absorption would occur in an available radio band, is low.}, a
significantly larger sample may not be available until the science operations of the Square Kilometre Array, or at least
its pathfinders (which are generally limited to $z_{\rm abs}\lapp1$, e.g. \citealt{asm+15a,mmo+17}).  In the meantime,
there are a further 176 intervening absorption systems which have been searched in \HI\ 21-cm absorption.  Adding these
to the sample increases its size by a factor of 3.5. In this paper, we explore the potential of using these systems to
provide a measure of the cold gas fraction and how this compares to the star formation history, with the view to future
surveys with the next generation of large radio telescopes.

\section{Analysis} 
\subsection{Line strengths of the intervening \HI\ 21-cm absorbers}

The total neutral atomic hydrogen column density, $N_{\rm HI}$ [\scm], is related to the velocity integrated optical
depth of the \HI\ 21-cm absorption via 
\begin{equation}
N_{\rm HI}  =1.823\times10^{18}\,T_{\rm  spin}\int\!\tau\,dv,
\label{enew_full}
\end{equation}
where the harmonic mean spin temperature, $T_{\rm spin}$, is a measure of the population of the lower hyperfine level ($F=1$), where
the gas can absorb 21-cm photons \citep{pf56}, relative to the upper hyperfine level ($F=2$). Comparison of the 21-cm line strength
with the total column density, from Lyman-\AL\ absorption along the same sight-line, therefore provides a thermometer,
where $T_{\rm spin} \propto N_{\text{\HI}}/\int\!\tau\,dv$. 

However, we cannot measure $\int\!\tau\,dv$ directly, since the observed optical depth, which is 
ratio of the line depth, $\Delta S$, to the observed background flux, $S_{\rm obs}$, is
related to the intrinsic optical depth via
\begin{equation}
\tau \equiv-\ln\left(1-\frac{\tau_{\rm obs}}{f}\right) \approx  \frac{\tau_{\rm obs}}{f}, {\rm ~for~}  \tau_{\rm obs}\equiv\frac{\Delta S}{S_{\rm obs}}\lapp0.3,
\label{tau_obs}
\end{equation}
where the covering factor, $f$, is the fraction of $S_{\rm obs}$ intercepted by the absorber.  Therefore, in the
optically thin regime (where $\tau_{\rm obs}\lapp0.3$), Equ. \ref{enew_full} can be approximated as
\begin{equation}
N_{\text{\HI}}  \approx 1.823\times10^{18}\,\frac{T_{\rm  spin}}{f}\int\!\tau_{\rm obs}\,dv.
\label{enew}
\end{equation}
So in order to measure the temperature, we require the velocity integrated optical depth of the 21-cm absorption profile
(as well as the total neutral hydrogen column density, discussed in Sect.~\ref{ngsf}).

For this study we compiled all of the published searches for redshifted intervening \HI\ 21-cm absorption towards
Quasi-Stellar Objects (QSOs).\footnote{\citet{dm78,bs79,bw83,cry93,lrj+96,cdn99,ck00,lan00,lb01,kcsp01,kgc01,kpec09,kem+12,kps+14,bdv01,kc01a,kc02,kb03,dgh+04,cmp+03,cdbw07,ctm+07,ctp+07,ctd+09,ykep07,gsp+09,gsp+09a,gsp+12,gsn+13,ekp+12,sgp+12,rmgn13,kan14,zlp+15,dsgj17,dsg+17}.}
This comprised 250 absorption systems, 74 of which have measured neutral hydrogen column densities
(i.e. DLAs or sub-DLAs), with the remaining 176 consisting of 167 Mg{\sc \,ii} absorbers and nine detected through other methods
(such as 21-cm spectral scans, \citealt{br73}). For each of these the observed parameters; velocity integrated optical
depth, r.m.s. noise limit, flux density at the redshifted 21-cm frequency\footnote{For \citet{lan00} the flux densities
  are not given and so we interpolated these from the neighbouring frequencies.}, full-width half maximum (FWHM) of
the profile and the observed spectral resolution were obtained from the compiled literature.
 
\begin{figure}
\centering \includegraphics[angle=-90,scale=0.52]{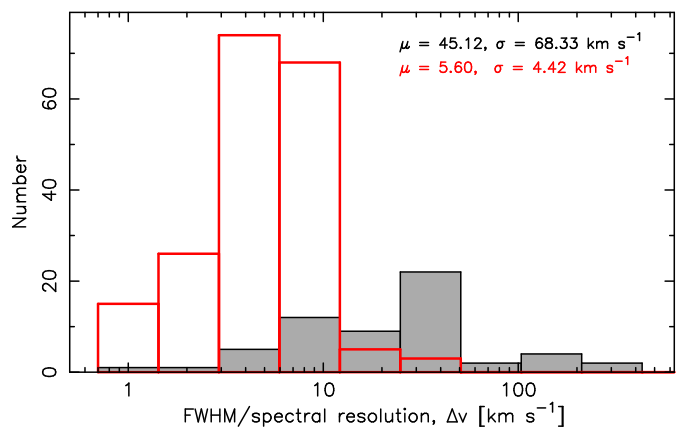}
\caption{The distribution of spectral resolution (for the non-detections, unfilled histogram) and the line-widths (for the
  detections, filled histogram).  The non-detections span a range of $0.29 - 30$ \kms, which are too disparate to show clearly on a linear scale.}
\label{vel-hist}
\end{figure}

As discussed in \citet{cur17}, in order to compare the 21-cm absorption results consistently, it is necessary to
normalise the sensitives. 
Since the spectral resolutions span a large range of values (Fig. \ref{vel-hist}),
we re-sample the r.m.s. noise levels to a common channel width, which is then used as FWHM of the putative absorption profile.
The detections have a mean profile width of $\left< {\rm FWHM}\right> = 45$ \kms, 
which we use to recalculate the $3\sigma$ upper limit to the integrated optical depth for each non-detection. i.e. 
$\int\tau_{\rm obs}dv  < 3 (\Delta S/S_{\rm obs}) \times 45$ \kms\ (cf. Equ.~\ref{tau_obs}).\footnote{This resampling results in a scaling of $\sqrt{{\rm FWHM}/\Delta v}$ to the r.m.s. noise level, where $\Delta v$ is the original resolution (\citealt{cur12}).}  
Since there is no evolution in the FWHM of the
intervening absorbers detected in 21-cm \citep{cdda16}, we do not consider any redshift dependence.

\begin{figure}
\centering \includegraphics[angle=-90,scale=0.54]{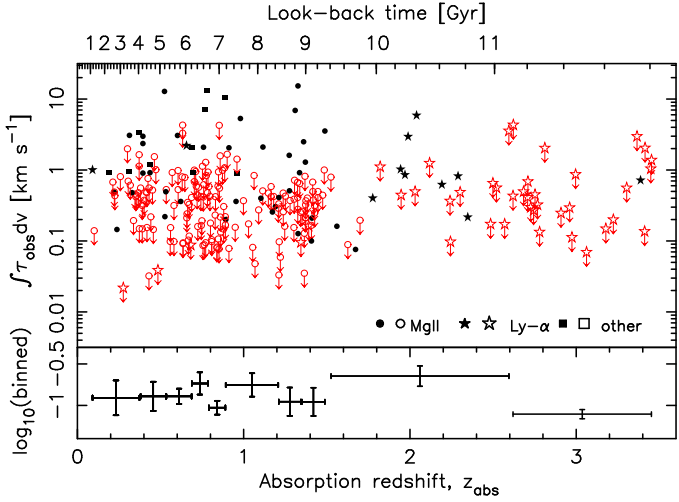}
\caption{The velocity integrated optical depth versus the redshift for the intervening absorbers searched in \HI\ 21-cm.
  The filled symbols show the detections and the unfilled circles the $3\sigma$ upper limits, with the shape
  representing the transition in which the absorption was initially detected: circles -- Mg{\sc ii}, stars -- Lyman-\AL,
  squares -- other (e.g. 21-cm scan).  The bottom panel shows the binned values in equally sized bins (10 bins of 25),
  including the limits, where the horizontal error bars show the range of points in the bin and the vertical error bars
  the $1\sigma$ uncertainty in the mean value.}
\label{int_tau-z}
\end{figure}

Using these and the values quoted in the literature for the detections, in Fig. \ref{int_tau-z} we show the
distribution of the velocity integrated optical depth of the absorption versus the redshift.
In the bottom panel, the upper limits are included in the binning as censored data points, via the {\em Astronomy
  SURVival Analysis} ({\sc asurv}) package \citep{ifn86}. The points are binned via the Kaplan--Meier estimator, giving the
maximum-likelihood estimate based upon the parent population \citep{fn85}, from which we see no overwhelming bias in the
survey sensitivity between the Mg{\sc ii} absorbers and DLAs.

\subsection{The spin temperature/covering factor degeneracy}
\label{stcf}

\begin{figure}
\centering \includegraphics[angle=-90,scale=0.54]{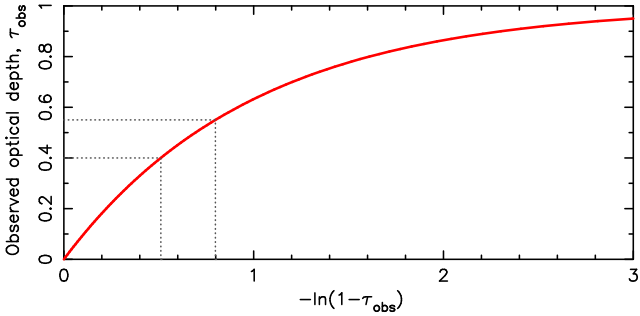}
\caption{The variation of the observed optical depth with the intrinsic optical depth. The dotted lines show the maximum effect
  (where $f=1$) on the two optically thick cases.}
\label{tau}
\end{figure}

Since only two of the detections exhibit optically thick ($\tau_{\rm obs}>0.3$) absorption, we can, in principle, use
Equ. \ref{enew} to determine the spin temperature of the gas. These two absorbers have peak optical depths of $\tau_{\rm
  obs}\approx0.40$ ($z_{\rm abs} = 1.3265$ towards J0850+5159) and $\tau_{\rm obs} \approx0.55$ ($z_{\rm abs} = 1.3603$
towards FBQS\,J2340--0053, \citealt{gsp+09a}), Fig. \ref{tau},
and so the range of possible intrinsic optical depths are $\tau = 0.40 - 0.51$ and $\tau = 0.55 -
0.80$, respectively (since $\tau_{\rm obs} <f \leq1$, \citealt{obg94}).

Since $T_{\rm spin} \propto f\,N_{\text{\HI}}/\int\!\tau_{\rm obs}\,dv$, we also require the covering factor to
determine the spin temperature. However, without knowledge of the relative extents of the absorber--frame 1420~MHz
absorption and emission cross-sections, nor the alignment between the absorber and the emitter, this is unknown.
It will, however, exhibit a strong redshift dependence \citep{cw06}: 
In the small angle approximation, this is given by 
\begin{equation}
f= \left\{   
\begin{array}{l l}
\left(\frac{d_{\rm abs} DA_{\rm QSO}}{d_{\rm QSO} DA_{ \rm abs}}\right)^2 & \text{ if } \theta_{\rm abs} < \theta_{\rm QSO}\\
  1  & \text{ if } \theta_{\rm abs} \geq\theta_{\rm QSO}\\
\end{array}
\right.  
\label{f}
\end{equation}
(see \citealt{cur12,azdc16}), where the angular diameter distance to a source is 
\begin{equation}
DA = \frac{DC}{z+1},{\rm ~where~} DC = \frac{c}{H_0}\int_{0}^{z}\frac{dz}{H_{\rm z}/H_0} 
\label{equ:DA}
\end{equation}
is the line-of-sight co-moving distance (e.g. \citealt{pea99}), 
in which $c$ is the speed of light, $H_0$ the Hubble constant and $H_{\rm z}$ the Hubble parameter at redshift $z$, given by 
${H_{\rm z}}/{H_{0}} = \sqrt{\Omega_{\rm m}\,(z+1)^3 + (1-\Omega_{\rm m} - \Omega_{\Lambda})\,(z+1)^2 + \Omega_{\Lambda}}$.
For a standard $\Lambda$ cosmology with $H_{0}=71$~km~s$^{-1}$~Mpc$^{-1}$, $\Omega_{\rm matter}=0.27$ and
$\Omega_{\Lambda}=0.73$, this gives a peak in the angular diameter distance at $z\approx1.6$, which has the consequence
that below this redshift $DA_{\rm DLA} \ll DA_{\rm QSO}$, as well as $DA_{\rm DLA} \sim DA_{\rm QSO}$, is possible (when $z_{\rm abs} \ll z_{\rm QSO}$),
whereas above $z_{\rm abs}\sim1.6$, {\em only} $DA_{\rm DLA} \sim DA_{\rm QSO}$ is possible. This
leads a mix of angular diameter distance ratios ($DA_{\rm abs}/DA_{ \rm QSO}$) at low redshift, but exclusively high
values ($DA_{\rm abs}/DA_{ \rm QSO}\sim1$) at high redshift.

\begin{figure}
\centering \includegraphics[angle=-90,scale=0.54]{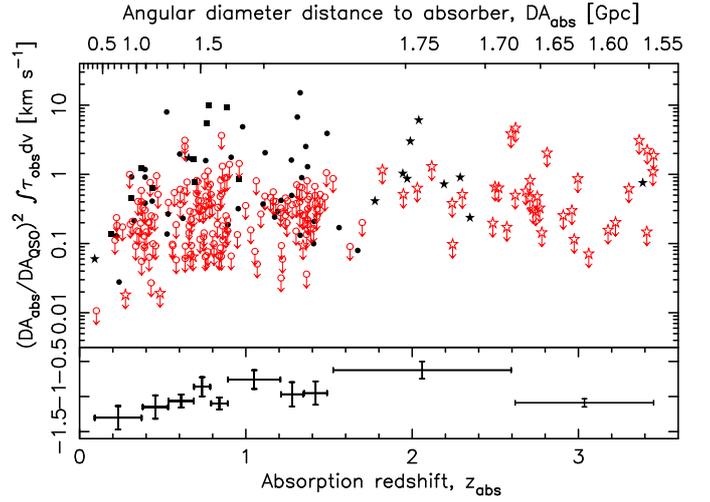}
\caption{As Fig. \ref{int_tau-z}, but corrected for the angular diameter distances (Equ.~\ref{equ:NoverT}).}
\label{int_tau-z_corr}
\end{figure}

Although we have no information on the absorber/emitter extents nor the DLA--QSO alignment, we can at least account for this
angular diameter bias, via (Equs. \ref{enew} \& \ref{f}, where $\theta_{\rm abs} < \theta_{\rm QSO}$)
\begin{equation}
\int\!\tau_{\rm obs}\,dv \left(\frac{DA_{\rm abs}}{DA_{\rm QSO}}\right)^2 = \frac{1}{1.823\times10^{18}}\frac{N_{\text{\HI}}}{T_{\rm spin}}\left(\frac{d_{\rm abs}}{d_{\rm QSO}}\right)^2,
\label{equ:NoverT}
\end{equation}
the effect of which we show in Fig. \ref{int_tau-z_corr}. 
From the binned data there may be a peak in $\int\!\tau_{\rm obs}\,dv \left({DA_{\rm abs}}/{DA_{\rm QSO}}\right)^2$ at
$z_{\rm abs}\sim2$, which is close to where the star formation density, $\psi_{*}$, peaks ($z=2.48$,
e.g. \citealt{hb06}).  Since the ordinate is proportional to $(N_{\text{\HI}}/T_{\rm spin})\left({d_{\rm abs}}/{d_{\rm
      QSO}}\right)^2$, this could indicate a physical connection, with the abundance of cool gas peaking close to the
maximum $\psi_{*}$, providing that there is no dominant evolution in ${d_{\rm abs}}/{d_{\rm QSO}}$. Without accounting
for the column density, however, this only demonstrates a peak in the abundance of cold gas, rather than in its
fraction.
  
\section{Evolution of the neutral gas}

\subsection{Neutral gas and star formation}
\label{ngsf}        

\begin{figure}
\centering \includegraphics[angle=-90,scale=0.54]{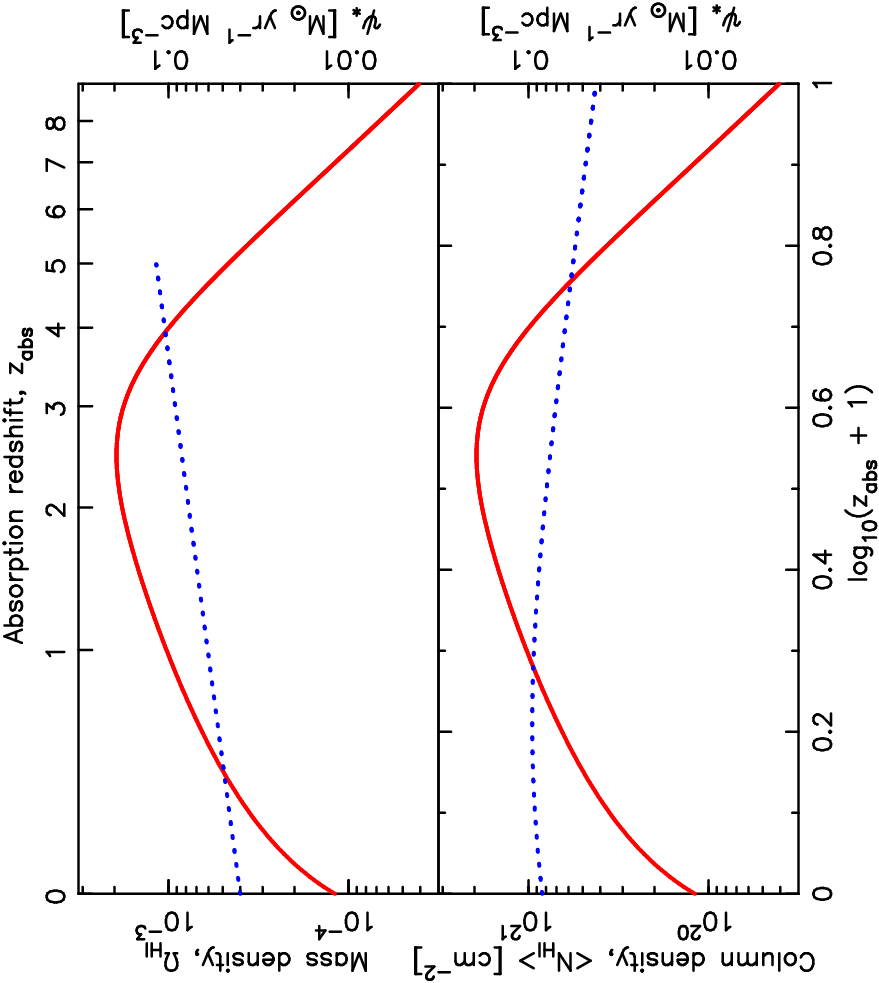}
\caption{The best fit to the cosmological \HI\ mass density of \citet{cmp+17} [$\Omega_{\text{\HI}} =
  4.0\times10^{-4}(z_{\rm abs}+1)^{0.60}$ -- dotted line, top panel] and the mean column density obtained
from this (bottom panel). The abscissa is mapped to $\log_{10}(z_{\rm abs} + 1)$, in order to demonstrate the
contrast between the evolution of $N_\text{\HI}$ and the star formation density \citep{hb06} [solid
  curve], where $\psi_{*}$ is arbitrarily shifted on the ordinate but retains the relative scaling (right hand scale).}
\label{N-SFR}
\end{figure}

Although for the Mg{\sc ii} absorbers we do not know the individual $N_{\text{\HI}}$ values, from the current 21-cm emission and
Lyman-\AL\ absorption data we do know how the mean column density, $\left< N_\text{\HI} \right>$, evolves with redshift. We can
obtain this from the evolution of the cosmological mass density (Fig. \ref{N-SFR}, top) via
\begin{equation}
\Omega_{\text{\HI}} = \frac{\mu\,m_{_{\rm H}}\,H_{0}}{c\,\rho_{\rm cr it}}\,n_{\rm DLA}\, \left< N_\text{\HI} \right>\frac{1}{(z+1)^2}\frac{H_{\rm z}}{H_{0}},
\end{equation} 
where $\mu = 1.3$ is a correction for the 75\% hydrogen composition, $m_{_{\rm H}}$ is the mass of the hydrogen atom,
$\rho_{\rm crit} \equiv 3\,H_{0}^2/8\,\pi\,G$ is the critical mass density of the Universe, where $G$ is the
gravitational constant, and $n_{\rm DLA} = 0.027(z_{\rm abs}+1)^{1.682}$ \citep{rtsm17} is the redshift number density of
DLAs.

\begin{figure}
\centering \includegraphics[angle=-90,scale=0.54]{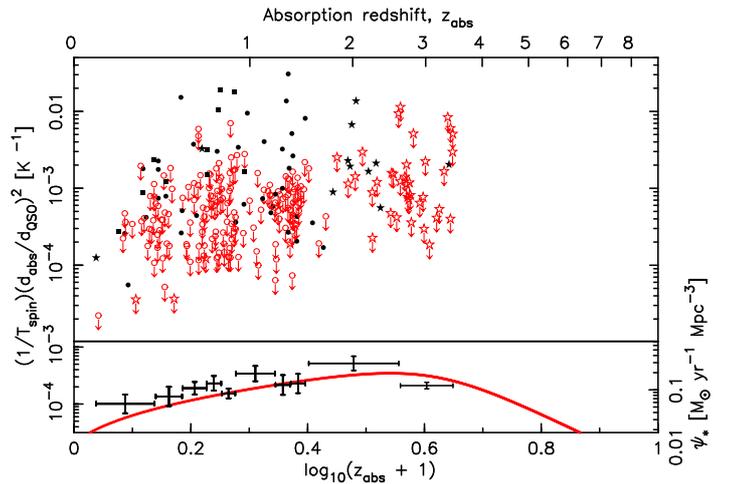}
\caption{The reciprocal of the spin temperature degenerate with the ratio of the absorber/emitter extents
  (Equ. \ref{equ:NoverT}). As per Fig. \ref{N-SFR}, the curve shows the best fit to the SFR density \citep{hb06}
arbitrarily shifted for comparison with the binned values of the top panel.}
\label{over_spin}
\end{figure}

\begin{figure}
\centering \includegraphics[angle=-90,scale=0.47]{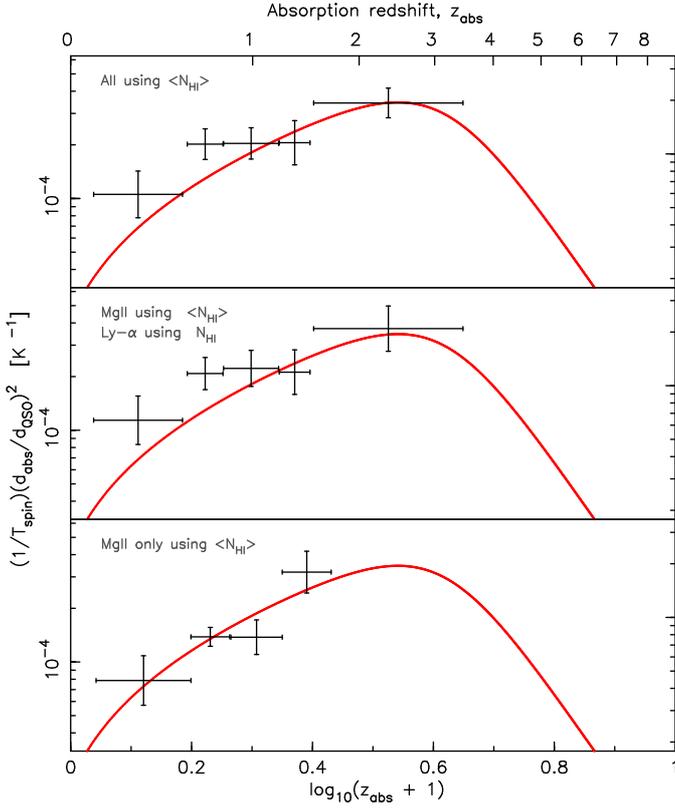}
\caption{As Fig. \ref{over_spin}, using the mean column density for all of the absorbers (top panel), only the Mg{\sc
    ii} absorbers normalised by $\left< N_\text{\HI} \right>$ with the DLAs normalised by the actual $N_\text{\HI}$
  measurements (see \citealt{cur17}, middle panel). In these two panels the bin size has been doubled to $n=50$ (halving
  the number of bins), in order to improve the signal-to-noise ratios. The bottom panel shows only the Mg{\sc ii}
  absorbers normalised by $\left< N_\text{\HI} \right>$, where the sample of 188 is binned into four bins of 47.  As per
  Fig. \ref{N-SFR}, the curves show the SFR density arbitrarily shifted on the ordinate, where the same shift is used
  in each panel.}
\label{3-spin}
\end{figure}

We show the derived distribution of $\left< N_\text{\HI} \right>$ in Fig.~\ref{N-SFR} (bottom) and applying this to
Equ.~\ref{equ:NoverT}, we can obtain the mean evolution in $({1}/{T_{\rm spin}})\left({d_{\rm abs}}/{d_{\rm QSO}}\right)^2$.
As per the DLAs, these appear to trace $\psi_{*}$, at least as far as the upper redshift limit of the absorbers
(Fig.~\ref{over_spin}).
Since actual column densities are available for the confirmed DLAs, which occupy the higher redshifts (Fig.~\ref{int_tau-z}), in Fig.~\ref{3-spin}
we also show the distribution using the measured column densities  for the DLAs, as well as applying $\left< N_\text{\HI} \right>$
to the Mg{\sc ii} absorbers alone.
From this, we see that actual column densities are consistent with the mean values, although the uncertainties are
larger  because of the smaller numbers. From the bottom panel, we see that $(1/T_{\rm spin})(d_{\rm
  abs}/d_{\rm QSO})^2$ for the Mg{\sc ii} absorbers only also traces $\psi_{*}$, although the redshift range is more
truncated (due to the $z_{\rm abs}\lapp2.2$ limitation of ground-based Mg{\sc ii} spectroscopy).

\begin{figure}
\centering \includegraphics[angle=-90,scale=0.56]{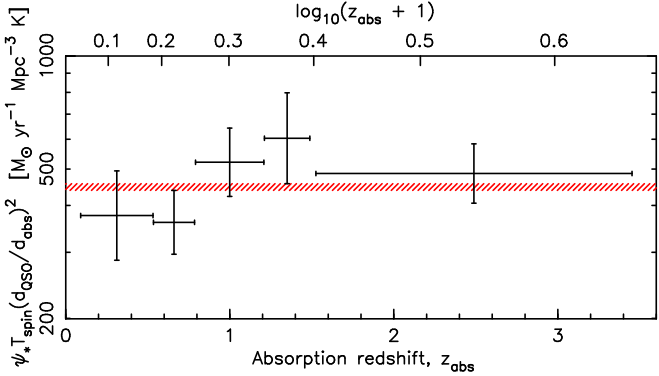}
\caption{The SFR density normalised by $({1}/{T_{\rm spin}})\left({d_{\rm abs}}/{d_{\rm QSO}}\right)^2$.
The hatching shows the region over which the error bars overlap, $439-457$~\Mo\ yr$^{-1}$ Mpc $^{-3}$ K.}
\label{SFRminusT}
\end{figure}

In order to test the similarity between $(1/T_{\rm spin})(d_{\rm abs}/d_{\rm QSO})^2$  and $\psi_{*}$, in Fig.~\ref{SFRminusT}
we show the SFR density normalised by the fraction of cool gas
\begin{equation}
\psi_{*}\left[\frac{1.823\times10^{18}}{\left<N_{\text{\HI}}\right>}\int\!\tau_{\rm obs}\,dv \left(\frac{DA_{\rm abs}}{DA_{\rm QSO}}\right)^2\right]^{-1} = \psi_{*}T_{\rm spin}\left(\frac{d_{\rm QSO}}{d_{\rm abs}}\right)^2,
\end{equation}
from which the residuals are consistent with zero redshift evolution, within the $\pm1\sigma$ uncertainties.
This  implies a direct correlation between these two quantities and the normalisation gives $\psi_{*}T_{\rm spin}\approx450 (d_{\rm abs}/d_{\rm
  QSO})^2$ ~\Mo~yr$^{-1}$~Mpc~$^{-3}$~K, for the dependence of the star formation density upon the spin temperature (see
Sect.~\ref{sfcnm}).

\subsection{Star formation and the fraction of cold neutral medium}
\label{sfcnm}

\begin{figure}
\centering \includegraphics[angle=-90,scale=0.54]{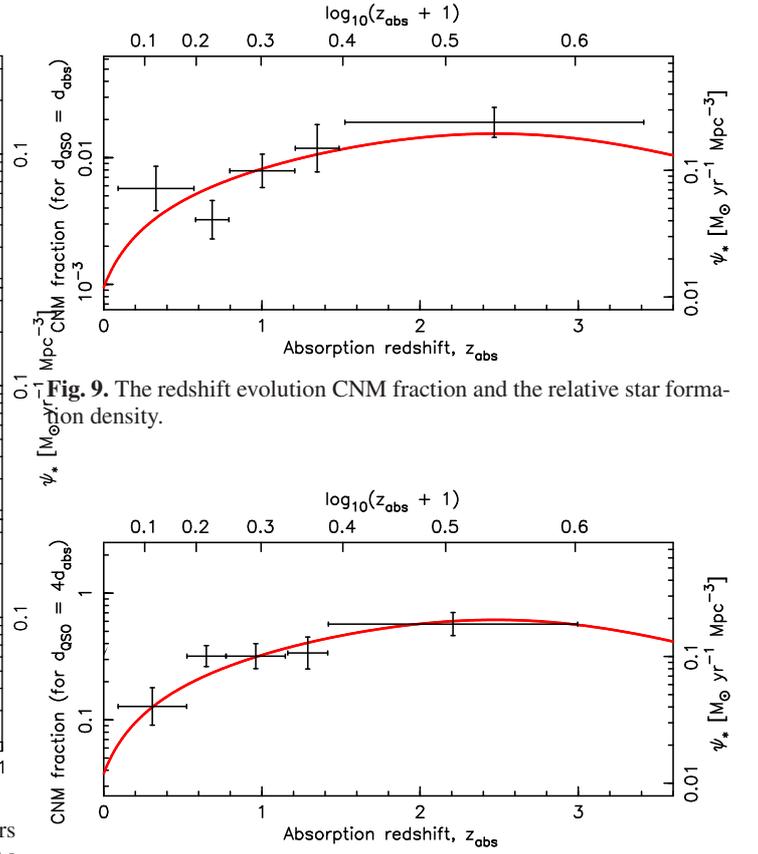}
\caption{The redshift evolution CNM fraction and the relative star formation density.}
\label{CNM}
\end{figure}

\begin{figure}
\centering \includegraphics[angle=-90,scale=0.54]{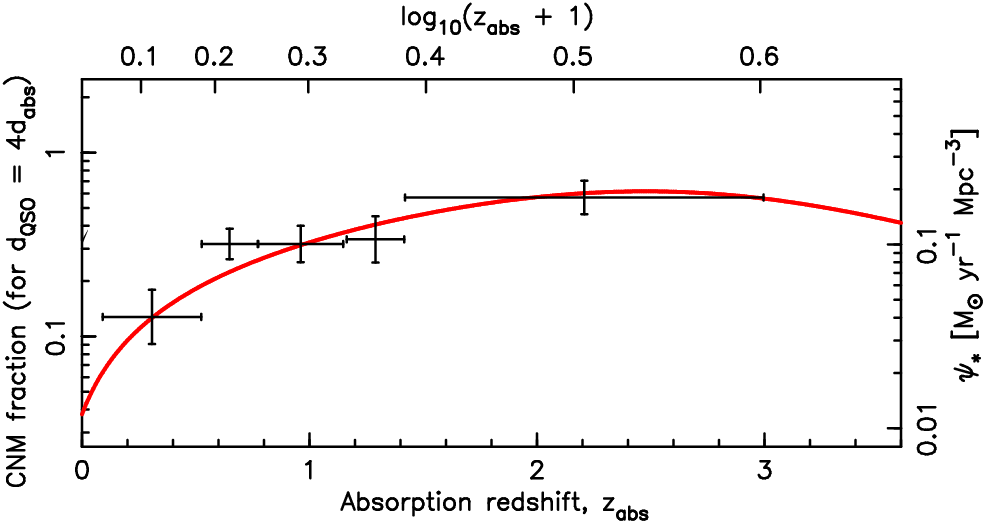}
\caption{As per Fig. \ref{CNM} but for $d_{\rm QSO} = 4d_{\rm abs}$.}
\label{CNM_scale}
\end{figure}

Neutral gas in the interstellar medium is hypothesised to comprise two components (\citealt{fgh69,whm+95}) -- the 
cold neutral medium (CNM, where $T\sim150$~K and $n\sim10$~\ccm)
and the warm neutral medium (WNM, where $T\sim10\,000$~K and $n\sim0.2$~\ccm). 
The CNM fraction is derived from the CNM, WNM and spin temperatures, via
\begin{equation}
{\cal F}_{\rm CNM} \equiv{\left[{\frac{1}{T_{\rm spin}} - \frac{1}{T_{\rm WNM}}}\right]} \Bigg/ {\left[{\frac{1}{T_{\rm CNM}} - \frac{1}{T_{\rm WNM}}}\right]},
\end{equation}
giving the distribution in Fig. \ref{CNM}.
Again, there is a reasonable trace of the star formation density, although the values are low compared to those
observed, specifically ${\cal F}_{\rm CNM}\approx0.3$ in the Milky Way \citep{ht03}, $z_{\rm abs} = 0.09$
\citep{lbs00} and $z_{\rm abs} = 0.22$ \citep{kgc01}, getting as high as ${\cal F}_{\rm CNM}\approx0.8$ at $z_{\rm
  abs} \approx2$ \citep{kps+14}. The spin temperature we derive is, however, degenerate with the ratio of the absorber/emitter
extents and gives CNM fractions similar to those observed if we apply $\left <d_{\rm QSO}\right> = 4\left<d_{\rm abs}\right>$
(Fig. \ref{CNM_scale}).
This ratio gives $\psi_{*}T_{\rm spin}\approx30$~\Mo~yr$^{-1}$~Mpc~$^{-3}$~K (cf. Fig.~\ref{SFRminusT}) and so for a
temperature of $T_{\rm spin}\approx300$~K, we may expect a star formation density of $\psi_{*}\approx0.1$ ~\Mo~yr$^{-1}$~Mpc~$^{-3}$. This is, of course, dependent on any evolution in $d_{\rm abs}/d_{\rm QSO}$, in addition to the assumption that the 
covering factor is generally less than unity (Equ. \ref{f}).

\begin{figure}
\centering \includegraphics[angle=-90,scale=0.54]{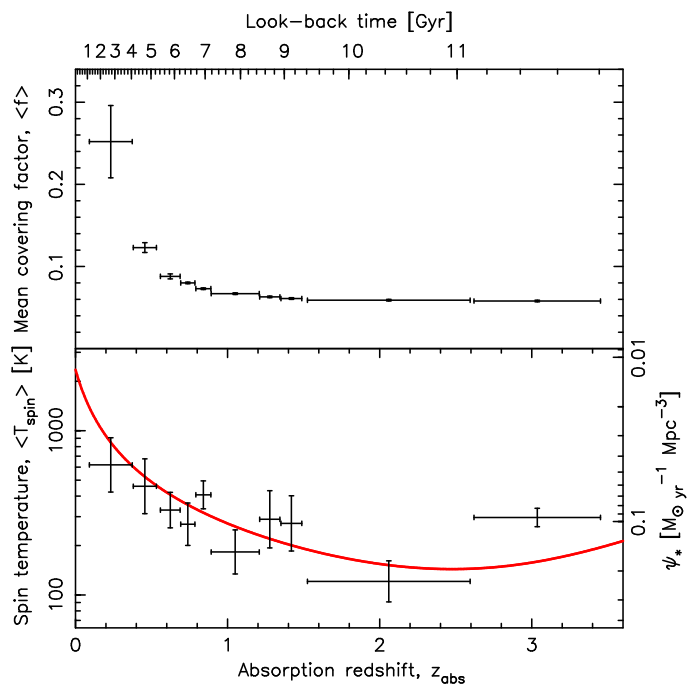}
\caption{The estimated covering factor and spin temperature evolution, assuming $f<1$ and $d_{\rm QSO} = 4d_{\rm abs}$. The curve shows $1/\psi_{*}$ \citep{hb06}, scaled according to  $\psi_{*}T_{\rm spin}\approx450 (d_{\rm abs}/d_{\rm QSO})^2$ ~\Mo\ yr$^{-1}$ Mpc $^{-3}$  (Sect. \ref{ngsf}).}
\label{f-T}
\end{figure}

With this assumption, $d_{\rm QSO}/d_{\rm abs}$ is the only unknown in Equ.~\ref{f} and so we can
use the estimate of the mean $d_{\rm QSO}/d_{\rm abs}$ ratio to determine the evolution of the mean covering factor from $DA_{\rm abs}
< DA_{\rm QSO}/4$, which applies to all of sample.\footnote{The largest ratio is from the $z_{\rm abs} = 0.091$ absorber
  towards the $z_{\rm QSO} = 0.635$ FBQS\,J074110.6+311200 \citep{lbs00}, which has $DA_{\rm abs}= 346$ Mpc and $DA_{\rm
    QSO} =1413$ Mpc, giving $DA_{\rm abs} =0.245 DA_{\rm QSO}$.}
The  mean covering factors derived are similar to those obtained from the Monte-Carlo simulation of \citet{cur17} and
the range of mean spin temperatures are consistent with those found in the Milky Way \citep{dsg+09} and other near-by
galaxies \citep{cras16}, $T_{\rm spin}\approx200 - 2000$~K  (Fig. \ref{f-T}).
We reiterate, however, that this assumes a mean $d_{\rm QSO} = 4d_{\rm abs}$ over all redshifts and a general covering
factor of less than unity.

\section{Possible caveats}
\label{pc}

\subsection{Column density estimates}
\label{cde}

One motivation for this work is to investigate the potential of using the evolution of the mean column density to obtain
$T_{\rm spin}/f$ from the 21-cm absorption strength, where individual column density measurements will not be practical.
For example, the 150\,000 sight-lines to be probed in the First Large Absorption Survey in \HI\ (FLASH) on the Australian SKA
Pathfinder (ASKAP, \citealt{asm+15a}).\footnote{Given the absence of an optical spectrum from which to determine the nature of the
  absorber, other techniques, such as machine learning, may be able to distinguish whether the absorption is intervening
  or associated with the background continuum source \citep{cdda16}.} Since no UV spectrometer is planned for the James Webb
Space Telescope, this will be a particular problem for the $z_{\text{\HI}}\lapp1$ limitation of the SKA pathfinders
(e.g. \citealt{mmo+17}) upon the demise of the Hubble Space Telescope.

\begin{figure}
\centering \includegraphics[angle=-90,scale=0.54]{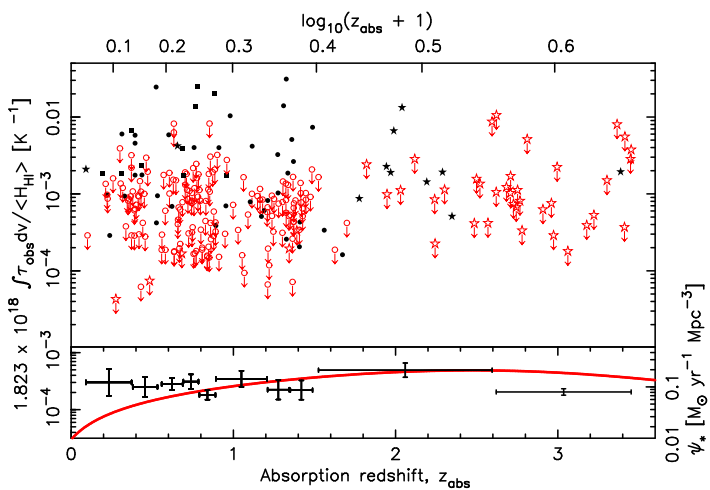}
\caption{The estimated covering factor/spin temperature degeneracy (uncorrected for geometry), $f/T_{\rm spin}$, obtained
  using $\left<N_\text{\HI} \right>$. }
\label{DLA-test}
\end{figure}

From the similarities between the distributions in Fig.~\ref{3-spin}, it does appear that the estimated column densities
are statistically consistent with the measured values. To test this, in Fig.~\ref{DLA-test} we show the distribution of
the 21-cm line strength normalised by the mean column density, $1.823\times10^{18}\int\!\tau_{\rm
  obs}\,dv/\left<N_\text{\HI} \right> = \left<f/T_{\rm spin}\right>$ (Equ.~\ref{enew}).
This bears a close resemblance to the $f/T_{\rm spin}$ distribution for DLAs \citep{cur17}, where there is also a
flattening of the distribution at low redshift.
\begin{figure}
\centering \includegraphics[angle=-90,scale=0.54]{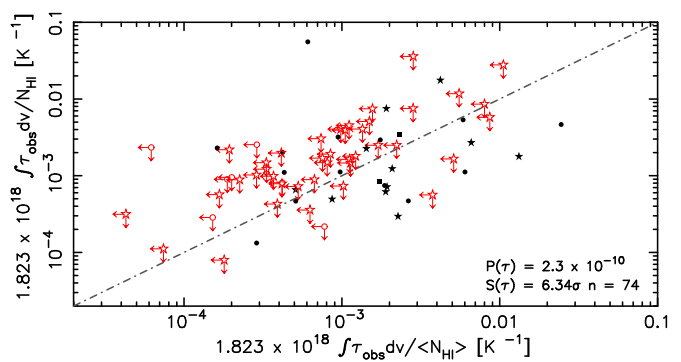}
\caption{The actual covering factor/spin temperature degeneracy (see \citealt{cur17}) in comparison to the estimated
  values (Fig.~\ref{DLA-test}).  The line has a gradient of unity and zero intercept.  A Kendall-tau test, including the
  limits, gives a probability $P(\tau) = 2.27\times10^{-10}$ of the observed distribution occuring by chance, which is
  significant at $S(\tau) = 6.34\sigma$, assuming Gaussian statistics.}
\label{N_comp}
\end{figure} 
In Fig.~\ref{N_comp}, we show the effect that the estimated column density has on the confirmed DLAs and sub-DLAs
searched in 21-cm absorption. Although there is  considerable spread, this small sample exhibits  a strong correlation.
This, and the similarity of Fig.~\ref{DLA-test} to the DLA distribution,
gives us confidence in the application of  this method to obtain a statistical estimate of the column density at a given redshift.

\subsection{The correction for geometry effects}
\label{ade}

\begin{figure}
\centering \includegraphics[angle=-90,scale=0.54]{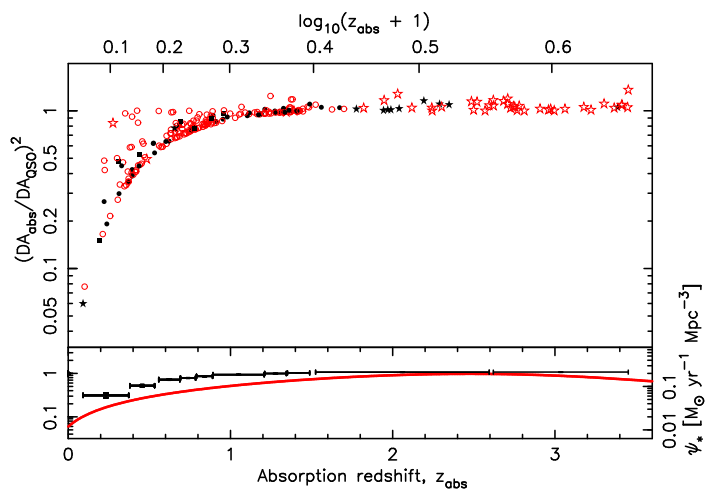}
\caption{The evolution of the angular diameter distance ratio with redshift. Again, the solid curve in the bottom
  panels shows the star formation density--- redshift distribution from \citet{hb06}.}
\label{ratio_lin}
\end{figure}

\begin{figure}
\centering \includegraphics[angle=-90,scale=0.54]{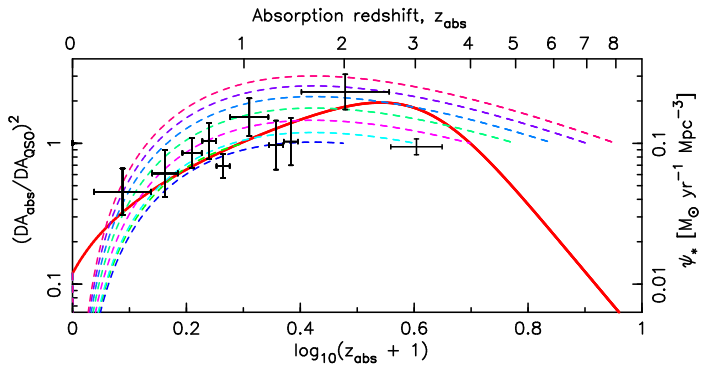}
\caption{The star formation evolution (solid curve) and $(1/T_{\rm spin})(d_{\rm abs}/d_{\rm QSO})^2$ (error bars,
  scaled by a factor of 450, Sect.~\ref{sfcnm}) superimposed upon the ratio of angular diameter distances for various
  QSO redshifts (broken curves). These are shown for $z_{\rm QSO}= 1, 2, ..., 8$ where the QSO redshift is given by the terminating
  value of the curve showing the absorption redshift distribution of $(DA_{\rm abs}/DA_{\rm QSO})^2$.}
\label{ratio_z}
\end{figure}

As previously stated, the above analysis assumes that there is no evolution in $d_{\rm abs}/d_{\rm QSO}$, in addition to
any absorber--emitter misalignment and emitter structure being averaged out. As well as this, the covering factors are
assumed to be generally less than unity. For $f<1$, $f\propto\left({DA_{\rm QSO}}/{DA_{\rm abs}}\right)^2$ [Equ. \ref{f}],
which we see, by the comparison of Figs. \ref{int_tau-z}  and \ref{int_tau-z_corr}, is the dominant effect in giving the similarity
in the redshift evolution (Fig.~\ref{DLA-test} cf. \ref{ratio_lin}).
Regarding this:
\begin{enumerate}
\item This implies that $\psi_{*}\propto \left({DA_{\rm abs}}/{DA_{\rm QSO}}\right)^2$. Since the latter is purely
  due to geometry,  there must be some more fundamental underlying parameter to which both parameters are
  related. This is most likely the redshift evolution which peaks at $z=2.5$, compared to $z_{\rm abs}=1.6$ for $DA_{\rm abs}/DA_{\rm QSO}$.
        \item Correcting the observed optical depth by the covering factor is necessary if $f<1$ (Sect. \ref{stcf}).
          Although we have no information on the relative sizes nor the alignment, we do know that the geometry of the expanding Universe
          introduces a systematic difference in the possible values of $DA_{\rm abs}/DA_{\rm QSO}$ between the low and high
           redshift regimes. Thus, this must be taken into account before 
           before making any comparison between the low and high redshift optical depths.
         \item If unjustified, adding the ``noise'' of the 21-cm absorption strength (Fig. \ref{DLA-test}), should not  
           improve the trace of the star formation density, otherwise this would be one coincidence on top of $DA_{\rm
             abs}/DA_{\rm QSO}$ exhibiting a similar evolution as $\psi_{*}$. In fact, although larger uncertainties are
           introduced, the product $\int\!\tau_{\rm obs}\,dv \left({DA_{\rm abs}}/{DA_{\rm QSO}}\right)^2$
           appears to ``reign in'' the outliers. Specifically, the systematic offset at $z_{\rm abs}\lapp2$ and the
           absence of a $z_{\rm abs}\gapp3$ downturn (Fig. \ref{ratio_lin} cf. Fig. \ref{over_spin}), also present in
           the uncorrected data (Fig. \ref{DLA-test}). Note that an increase
           in the spin temperature at these redshifts is also advocated  by \citet{rmgn13} and \citet{kps+14}.
\end{enumerate}
Provided that the assumptions are reasonable, the spin temperature shows a very similar evolution to the 
star formation density, which is diluted out by a similar evolution in the covering factor 
(Fig. \ref{f-T}), resulting in a flat distribution of $\int\!\tau_{\rm obs}\,dv \approx f\int\!\tau\,dv \propto N_{\text{\HI}}/T_{\rm spin}$ (Fig. \ref{DLA-test}).
 While further 21-cm observations of absorbers at low redshift will reduce the uncertainties introduced by
$\int\!\tau_{\rm obs}\,dv$, observations at high redshift could be conclusive in determining whether $\int\!\tau_{\rm
  obs}\,dv \left({DA_{\rm abs}}/{DA_{\rm QSO}}\right)^2$ follows the downturn traced by $\psi_{*}$ (Fig.~\ref{ratio_z}). 
As it stands,  the top bin is consistent with the ratio of angular diameter distances for $z_{\rm QSO}\approx3-4$ and
$z_{\rm abs}\gapp2.5$, although we reiterate that a correction for the angular diameter distances is required in order
to combine the low and high redshift populations.  From the figure it is clear that further high redshift data, particularly
at $z_{\rm abs}\gapp4$, would be conclusive.

\section{Summary}

It is an outstanding problem that the total neutral gas content of intervening absorbers does not trace the star
formation density, $\psi_{*}$, which shows strong evolution with redshift. Recently, however, by using the ratio of the
strength of the 21-cm absorption to the total column density as a thermometer, \citet{cur17} showed that the cool
component of the gas could trace $\psi_{*}$.  This is physically motivated, since the star formation requires that the
gas be sufficiently cool for the cloud to collapse under its own gravity, this cool gas usually being evident through
large molecular abundances in the ``giant molecular clouds'' which host the cool gas. Indeed there may be similar
correlation between the \MOLH\ density and $\psi_{*}$, at least up to $z\sim2$ (\citealt{lbz+14} and references
therein).

The \HI\  data are, however, limited to a sample of 74 absorbers where both  21-cm absorption has been searched and the column
density is known. 
By binning these in order to overcome individual line-of-sight effects, such as the absorber--QSO alignment, 
structure in the radio emission and situations where the covering factor may be unity, gives just three bins, which 
exhibit the same approximate peak at a similar relative magnitude as the star formation density \citep{cur17}.
Since 21-cm absorption searches of a significantly larger sample of DLAs will most likely have to wait for the 
Square kilometre Array, we  examine the potential of using other intervening absorption systems, where the neutral hydrogen column density
is not readily available (e.g. Mg{\sc \,ii} absorbers), to trace the star formation history. In order to do this, we:
\begin{enumerate}
  \item Normalise the upper limits in the integrated optical depth to the same spectral resolution and include these via 
    a survival analysis, giving a sample total of 250 absorbers.

  \item Remove the bias introduced to the covering factor between the low and high redshift absorbers by the geometry
    effects of an expanding Universe.  That is, correcting for the fact that absorbers at $z_{\rm abs}\gapp1.6$ are
    always at a similar angular diameter distance as the background continuum source, whereas at lower redshift there is
    a mix of angular diameter distances.

\item Assign a column density derived from the evolution of the cosmological \HI\ density.

\item Bin the data in order to average out differences in the individual line-of-sight effects.
\end{enumerate}
This yields $(1/T_{\rm spin})(d_{\rm abs}/d_{\rm QSO})^2$, which, as for the DLAs, appears to trace the 
star formation density. For no evolution in the ratio of the absorber--QSO sizes, this would imply that
$\psi_{*}\propto 1/T_{\rm spin}$.

As is the case for the DLA-only sample, however, data is lacking at higher redshifts ($z_{\rm abs}\gapp3$), meaning that
we cannot be certain that $(1/T_{\rm spin})(d_{\rm abs}/d_{\rm QSO})^2$ follows the same downturn as $\psi_{*}$ at
look-back times beyond 11 Gyr.  However, given that $\left< N_\text{\HI} \right>$ up to $z\sim5$ is known, we may need
only search for intervening 21-cm absorption.  A non-reliance upon an optical spectrum would be advantageous to the next
generation of radio band surveys, since optically selected surveys may miss the most dust reddened objects
\citep{wfp+95,cmr+98,cwa+17}. This does, however, depend upon the evolution in $\left< N_\text{\HI} \right>$ being
applicable to the optically faint objects. In any case, follow-up observations of the many newly discovered  21-cm
absorbers expected with the Square Kilometre Array \citep{msc+15}, with either 21-cm emission (limited to
$z_{\rm abs}\lapp1$, \citealt{so15}) or Lyman-\AL\ absorption (limited to $z_{\rm abs}\gapp1.7$), will be very
observationally expensive.  Upon confirmation with further data, it is hoped that the methods presented here may offer a
solution in determining the evolution of the cold gas fraction over large look-back times.

\section*{Acknowledgements}

I would like to thank the anonymous referee for their helpful comments, as well as James Allison for useful comments on
a draft of the manuscript.  This research has made use of the NASA/IPAC Extragalactic Database (NED) which is operated
by the Jet Propulsion Laboratory, California Institute of Technology, under contract with the National Aeronautics and
Space Administration and NASA's Astrophysics Data System Bibliographic Service. This research has also made use of
NASA's Astrophysics Data System Bibliographic Service and {\sc asurv} Rev 1.2 \citep{lif92a}, which implements the
methods presented in \citet{ifn86}.


\begin{thebibliography}{93}
\expandafter\ifx\csname natexlab\endcsname\relax\def\natexlab#1{#1}\fi

\bibitem[{{Allison} {et~al.}(2017){Allison}, {Moss}, {Macquart}, {Curran},
  {Duchesne}, {Mahony}, {Sadler}, {Whiting}, {Bannister}, {Chippendale},
  {Edwards}, {Harvey-Smith}, {Heywood}, {Indermuehle}, {Lenc}, {Marvil},
  {McConnell}, \& {Sault}}]{amm+16}
{Allison}, J.~R., {Moss}, V.~A., {Macquart}, J.-P., {et~al.} 2017, MNRAS, 465,
  4450

\bibitem[{{Allison} {et~al.}(2016{\natexlab{a}}){Allison}, {Sadler}, {Moss},
  {Harvey-Smith}, {Heywood}, {Indermuehle}, {McConnell}, {Sault}, \&
  {Whiting}}]{asm+15a}
{Allison}, J.~R., {Sadler}, E.~M., {Moss}, V.~A., {et~al.} 2016{\natexlab{a}},
  Astronomische Nachrichten, 337, 175

\bibitem[{{Allison} {et~al.}(2016{\natexlab{b}}){Allison}, {Zwaan}, {Duchesne},
  \& {Curran}}]{azdc16}
{Allison}, J.~R., {Zwaan}, M.~A., {Duchesne}, S.~W., \& {Curran}, S.~J.
  2016{\natexlab{b}}, MNRAS, 462, 1341

\bibitem[{{Braun}(2012)}]{bra12}
{Braun}, R. 2012, ApJ, 87, 749

\bibitem[{{Briggs} {et~al.}(2001){Briggs}, {de Bruyn}, \& {Vermeulen}}]{bdv01}
{Briggs}, F.~H., {de Bruyn}, A.~G., \& {Vermeulen}, R.~C. 2001, A\&A, 373, 113

\bibitem[{Briggs \& Wolfe(1983)}]{bw83}
Briggs, F.~H. \& Wolfe, A.~M. 1983, ApJ, 268, 76

\bibitem[{{Brown} \& {Roberts}(1973)}]{br73}
{Brown}, R.~L. \& {Roberts}, M.~S. 1973, ApJ, 184, L7

\bibitem[{{Brown} \& {Spencer}(1979)}]{bs79}
{Brown}, R.~L. \& {Spencer}, R.~E. 1979, ApJ, 230, L1

\bibitem[{{Burgarella} {et~al.}(2013){Burgarella}, {Buat}, {Gruppioni},
  {Cucciati}, {Heinis}, {Berta}, {B{\'e}thermin}, {Bock}, {Cooray}, {Dunlop},
  {Farrah}, {Franceschini}, {Le Floc'h}, {Lutz}, {Magnelli}, {Nordon},
  {Oliver}, {Page}, {Popesso}, {Pozzi}, {Riguccini}, {Vaccari}, \&
  {Viero}}]{bbg+13}
{Burgarella}, D., {Buat}, V., {Gruppioni}, C., {et~al.} 2013, A\&A, 554, A70

\bibitem[{{Carilli} {et~al.}(1998){Carilli}, {Menten}, {Reid}, {Rupen}, \&
  {Yun}}]{cmr+98}
{Carilli}, C.~L., {Menten}, K.~M., {Reid}, M.~J., {Rupen}, M.~P., \& {Yun},
  M.~S. 1998, ApJ, 494, 175

\bibitem[{{Carilli} {et~al.}(1993){Carilli}, {Rupen}, \& {Yanny}}]{cry93}
{Carilli}, C.~L., {Rupen}, M.~P., \& {Yanny}, B. 1993, ApJ, 412, L59

\bibitem[{{Carilli} \& {Walter}(2013)}]{cw13}
{Carilli}, C.~L. \& {Walter}, F. 2013, ARA\&A, 51, 105

\bibitem[{{Chengalur} {et~al.}(1999){Chengalur}, {de Bruyn}, \&
  {Narasimha}}]{cdn99}
{Chengalur}, J.~N., {de Bruyn}, A.~G., \& {Narasimha}, D. 1999, A\&A, 343, L79

\bibitem[{{Chengalur} \& {Kanekar}(2000)}]{ck00}
{Chengalur}, J.~N. \& {Kanekar}, N. 2000, MNRAS, 318, 303

\bibitem[{{Crighton} {et~al.}(2015){Crighton}, {Murphy}, {Prochaska},
  {Worseck}, {Rafelski}, {Becker}, {Ellison}, {Fumagalli}, {Lopez}, {Meiksin},
  \& {O'Meara}}]{cmp+15}
{Crighton}, N.~H.~M., {Murphy}, M.~T., {Prochaska}, J.~X., {et~al.} 2015,
  MNRAS, 452, 217

\bibitem[{{Crighton} {et~al.}(2017){Crighton}, {Murphy}, {Prochaska},
  {Worseck}, {Rafelski}, {Becker}, {Ellison}, {Fumagalli}, {Lopez}, {Meiksin},
  \& {O'Meara}}]{cmp+17}
{Crighton}, N.~H.~M., {Murphy}, M.~T., {Prochaska}, J.~X., {et~al.} 2017, in
  IAU Symposium, Vol. 321, Formation and Evolution of Galaxy Outskirts, ed.
  A.~{Gil de Paz}, J.~H. {Knapen}, \& J.~C. {Lee}, 309--314

\bibitem[{Curran(2010)}]{cur09a}
Curran, S.~J. 2010, MNRAS, 402, 2657

\bibitem[{Curran(2012)}]{cur12}
Curran, S.~J. 2012, ApJ, 748, L18

\bibitem[{Curran(2017)}]{cur17}
Curran, S.~J. 2017, MNRAS, 470, 3159

\bibitem[{Curran {et~al.}(2007{\natexlab{a}})Curran, Darling, Bolatto, Whiting,
  Bignell, \& Webb}]{cdbw07}
Curran, S.~J., Darling, J.~K., Bolatto, A.~D., {et~al.} 2007{\natexlab{a}},
  MNRAS, 382, L11

\bibitem[{{Curran} {et~al.}(2016{\natexlab{a}}){Curran}, {Duchesne}, {Divoli},
  \& {Allison}}]{cdda16}
{Curran}, S.~J., {Duchesne}, S.~W., {Divoli}, A., \& {Allison}, J.~R.
  2016{\natexlab{a}}, MNRAS, 462, 4197

\bibitem[{Curran {et~al.}(2004)Curran, Murphy, Pihlstr\"{o}m, Webb, Bolatto, \&
  Bower}]{cmpw03}
Curran, S.~J., Murphy, M.~T., Pihlstr\"{o}m, Y.~M., {et~al.} 2004, MNRAS, 352,
  563

\bibitem[{Curran {et~al.}(2005)Curran, Murphy, Pihlstr\"{o}m, Webb, \&
  Purcell}]{cmp+03}
Curran, S.~J., Murphy, M.~T., Pihlstr\"{o}m, Y.~M., Webb, J.~K., \& Purcell,
  C.~R. 2005, MNRAS, 356, 1509

\bibitem[{{Curran} {et~al.}(2016{\natexlab{b}}){Curran}, {Reeves}, {Allison},
  \& {Sadler}}]{cras16}
{Curran}, S.~J., {Reeves}, S.~N., {Allison}, J.~R., \& {Sadler}, E.~M.
  2016{\natexlab{b}}, MNRAS, 459, 4136

\bibitem[{Curran {et~al.}(2010)Curran, Tzanavaris, Darling, Whiting, Webb,
  Bignell, Athreya, \& Murphy}]{ctd+09}
Curran, S.~J., Tzanavaris, P., Darling, J.~K., {et~al.} 2010, MNRAS, 402, 35

\bibitem[{Curran {et~al.}(2007{\natexlab{b}})Curran, Tzanavaris, Murphy, Webb,
  \& Pihlstr\"{o}m}]{ctm+07}
Curran, S.~J., Tzanavaris, P., Murphy, M.~T., Webb, J.~K., \& Pihlstr\"{o}m,
  Y.~M. 2007{\natexlab{b}}, MNRAS, 381, L6

\bibitem[{Curran {et~al.}(2007{\natexlab{c}})Curran, Tzanavaris, Pihlstr\"{o}m,
  \& Webb}]{ctp+07}
Curran, S.~J., Tzanavaris, P., Pihlstr\"{o}m, Y.~M., \& Webb, J.~K.
  2007{\natexlab{c}}, MNRAS, 382, 1331

\bibitem[{Curran \& Webb(2006)}]{cw06}
Curran, S.~J. \& Webb, J.~K. 2006, MNRAS, 371, 356

\bibitem[{Curran {et~al.}(2002)Curran, Webb, Murphy, Bandiera, Corbelli, \&
  Flambaum}]{cwbc01}
Curran, S.~J., Webb, J.~K., Murphy, M.~T., {et~al.} 2002, PASA, 19, 455

\bibitem[{{Curran} {et~al.}(2017){Curran}, {Whiting}, {Allison}, {Tanna},
  {Sadler}, \& {Athreya}}]{cwa+17}
{Curran}, S.~J., {Whiting}, M.~T., {Allison}, J.~R., {et~al.} 2017, MNRAS, 467,
  4514

\bibitem[{{Curran} {et~al.}(2006){Curran}, {Whiting}, {Murphy}, {Webb},
  {Longmore}, {Pihlstr{\"o}m}, {Athreya}, \& {Blake}}]{cwm+06}
{Curran}, S.~J., {Whiting}, M.~T., {Murphy}, M.~T., {et~al.} 2006, MNRAS, 371,
  431

\bibitem[{{Darling} {et~al.}(2004){Darling}, {Giovanelli}, {Haynes}, {Bower},
  \& {Bolatto}}]{dgh+04}
{Darling}, J., {Giovanelli}, R., {Haynes}, M.~P., {Bower}, G.~C., \& {Bolatto},
  A.~D. 2004, ApJ, 613, L101

\bibitem[{{Davis} \& {May}(1978)}]{dm78}
{Davis}, M.~M. \& {May}, L.~S. 1978, ApJ, 219, 1

\bibitem[{{Delhaize} {et~al.}(2013){Delhaize}, {Meyer}, {Staveley-Smith}, \&
  {Boyle}}]{dsmb13}
{Delhaize}, J., {Meyer}, M.~J., {Staveley-Smith}, L., \& {Boyle}, B.~J. 2013,
  MNRAS, 433, 1398

\bibitem[{{Dickey} {et~al.}(2009){Dickey}, {Strasser}, {Gaensler}, {Haverkorn},
  {Kavars}, {McClure-Griffiths}, {Stil}, \& {Taylor}}]{dsg+09}
{Dickey}, J.~M., {Strasser}, S., {Gaensler}, B.~M., {et~al.} 2009, ApJ, 693,
  1250

\bibitem[{{Dutta} {et~al.}(2017{\natexlab{a}}){Dutta}, {Srianand}, {Gupta}, \&
  {Joshi}}]{dsgj17}
{Dutta}, R., {Srianand}, R., {Gupta}, N., \& {Joshi}, R. 2017{\natexlab{a}},
  MNRAS, 468, 1029

\bibitem[{{Dutta} {et~al.}(2017{\natexlab{b}}){Dutta}, {Srianand}, {Gupta},
  {Joshi}, {Petitjean}, {Noterdaeme}, {Ge}, \& {Krogager}}]{dsg+17}
{Dutta}, R., {Srianand}, R., {Gupta}, N., {et~al.} 2017{\natexlab{b}}, MNRAS,
  465, 4249

\bibitem[{Ellison {et~al.}(2012)Ellison, Kanekar, Momjian, \& Worseck}]{ekp+12}
Ellison, S., Kanekar, N. amd~Prochaska, J.~X., Momjian, E., \& Worseck, G.
  2012, MNRAS, 424, 293

\bibitem[{{Feigelson} \& {Nelson}(1985)}]{fn85}
{Feigelson}, E.~D. \& {Nelson}, P.~I. 1985, ApJ, 293, 192

\bibitem[{{Fern{\'a}ndez} {et~al.}(2016){Fern{\'a}ndez}, {Gim}, {van Gorkom},
  {Yun}, {Momjian}, {Popping}, {Chomiuk}, {Hess}, {Hunt}, {Kreckel}, {Lucero},
  {Maddox}, {Oosterloo}, {Pisano}, {Verheijen}, {Hales}, {Chung}, {Dodson},
  {Golap}, {Gross}, {Henning}, {Hibbard}, {Jaff{\'e}}, {Donovan Meyer},
  {Meyer}, {Sanchez-Barrantes}, {Schiminovich}, {Wicenec}, {Wilcots},
  {Bershady}, {Scoville}, {Strader}, {Tremou}, {Salinas}, \&
  {Ch{\'a}vez}}]{fgv+16}
{Fern{\'a}ndez}, X., {Gim}, H.~B., {van Gorkom}, J.~H., {et~al.} 2016, ApJ,
  824, L1

\bibitem[{{Field} {et~al.}(1969){Field}, {Goldsmith}, \& {Habing}}]{fgh69}
{Field}, G.~B., {Goldsmith}, D.~W., \& {Habing}, H.~J. 1969, ApJ, 155, L149

\bibitem[{{Gupta} {et~al.}(2013){Gupta}, {Srianand}, {Noterdaeme}, {Petitjean},
  \& {Muzahid}}]{gsn+13}
{Gupta}, N., {Srianand}, R., {Noterdaeme}, P., {Petitjean}, P., \& {Muzahid},
  S. 2013, A\&A, 558, A84

\bibitem[{{Gupta} {et~al.}(2012){Gupta}, {Srianand}, {Petitjean}, {Bergeron},
  {Noterdaeme}, \& {Muzahid}}]{gsp+12}
{Gupta}, N., {Srianand}, R., {Petitjean}, P., {et~al.} 2012, A\&A, 544, 21

\bibitem[{{Gupta} {et~al.}(2009{\natexlab{a}}){Gupta}, {Srianand}, {Petitjean},
  {Noterdaeme}, \& {Saikia}}]{gsp+09}
{Gupta}, N., {Srianand}, R., {Petitjean}, P., {Noterdaeme}, P., \& {Saikia},
  D.~J. 2009{\natexlab{a}}, in Astronomical Society of the Pacific Conference
  Series, Vol. 407, The Low-Frequency Radio Universe, ed. D.~J. {Saikia}, D.~A.
  {Green}, Y.~{Gupta}, \& T.~{Venturi}, 67

\bibitem[{{Gupta} {et~al.}(2009{\natexlab{b}}){Gupta}, {Srianand}, {Petitjean},
  {Noterdaeme}, \& {Saikia}}]{gsp+09a}
{Gupta}, N., {Srianand}, R., {Petitjean}, P., {Noterdaeme}, P., \& {Saikia},
  D.~J. 2009{\natexlab{b}}, MNRAS, 398, 201

\bibitem[{{Heiles} \& {Troland}(2003)}]{ht03}
{Heiles}, C. \& {Troland}, T.~H. 2003, ApJ, 586, 1067

\bibitem[{{Hopkins} \& {Beacom}(2006)}]{hb06}
{Hopkins}, A.~M. \& {Beacom}, J.~F. 2006, ApJ, 651, 142

\bibitem[{{Hoppmann} {et~al.}(2015){Hoppmann}, {Staveley-Smith}, {Freudling},
  {Zwaan}, {Minchin}, \& {Calabretta}}]{hsf+15}
{Hoppmann}, L., {Staveley-Smith}, L., {Freudling}, W., {et~al.} 2015, MNRAS,
  452, 3726

\bibitem[{{Isobe} {et~al.}(1986){Isobe}, {Feigelson}, \& {Nelson}}]{ifn86}
{Isobe}, T., {Feigelson}, E., \& {Nelson}, P. 1986, ApJ, 306, 490

\bibitem[{{Kanekar}(2014)}]{kan14}
{Kanekar}, N. 2014, ApJ, 797, L20

\bibitem[{Kanekar \& Briggs(2003)}]{kb03}
Kanekar, N. \& Briggs, F.~H. 2003, A\&A, 412, L29

\bibitem[{Kanekar \& Chengalur(2001)}]{kc01a}
Kanekar, N. \& Chengalur, J.~N. 2001, A\&A, 369, 42

\bibitem[{Kanekar \& Chengalur(2003)}]{kc02}
Kanekar, N. \& Chengalur, J.~N. 2003, A\&A, 399, 857

\bibitem[{{Kanekar} {et~al.}(2001{\natexlab{a}}){Kanekar}, {Chengalur},
  {Subrahmanyan}, \& {Petitjean}}]{kcsp01}
{Kanekar}, N., {Chengalur}, J.~N., {Subrahmanyan}, R., \& {Petitjean}, P.
  2001{\natexlab{a}}, A\&A, 367, 46

\bibitem[{{Kanekar} {et~al.}(2013){Kanekar}, {Ellison}, {Momjian}, {York}, \&
  {Pettini}}]{kem+12}
{Kanekar}, N., {Ellison}, S.~L., {Momjian}, E., {York}, B.~A., \& {Pettini}, M.
  2013, MNRAS, 532

\bibitem[{{Kanekar} {et~al.}(2001{\natexlab{b}}){Kanekar}, {Ghosh}, \&
  {Chengalur}}]{kgc01}
{Kanekar}, N., {Ghosh}, T., \& {Chengalur}, J.~N. 2001{\natexlab{b}}, A\&A,
  373, 394

\bibitem[{{Kanekar} {et~al.}(2009){Kanekar}, {Prochaska}, {Ellison}, \&
  {Chengalur}}]{kpec09}
{Kanekar}, N., {Prochaska}, J.~X., {Ellison}, S.~L., \& {Chengalur}, J.~N.
  2009, MNRAS, 396, 385

\bibitem[{{Kanekar} {et~al.}(2014){Kanekar}, {Prochaska}, {Smette}, {Ellison},
  {Ryan-Weber}, {Momjian}, {Briggs}, {Lane}, {Chengalur}, {Delafosse}, {Grave},
  {Jacobsen}, \& {de Bruyn}}]{kps+14}
{Kanekar}, N., {Prochaska}, J.~X., {Smette}, A., {et~al.} 2014, MNRAS, 438,
  2131

\bibitem[{{Lagos} {et~al.}(2014){Lagos}, {Baugh}, {Zwaan}, {Lacey},
  {Gonzalez-Perez}, {Power}, {Swinbank}, \& {van Kampen}}]{lbz+14}
{Lagos}, C.~D.~P., {Baugh}, C.~M., {Zwaan}, M.~A., {et~al.} 2014, MNRAS, 440,
  920

\bibitem[{{Lah} {et~al.}(2007){Lah}, {Chengalur}, {Briggs}, {Colless}, {de
  Propris}, {Pracy}, {de Blok}, {Fujita}, {Ajiki}, {Shioya}, {Nagao},
  {Murayama}, {Taniguchi}, {Yagi}, \& {Okamura}}]{lcb+07}
{Lah}, P., {Chengalur}, J.~N., {Briggs}, F.~H., {et~al.} 2007, MNRAS, 376, 1357

\bibitem[{Lane(2000)}]{lan00}
Lane, W.~M. 2000, PhD thesis, University of Groningen

\bibitem[{Lane \& Briggs(2001)}]{lb01}
Lane, W.~M. \& Briggs, F.~H. 2001, ApJ, 561, L27

\bibitem[{{Lane} {et~al.}(2000){Lane}, {Briggs}, \& {Smette}}]{lbs00}
{Lane}, W.~M., {Briggs}, F.~H., \& {Smette}, A. 2000, ApJ, 532, 146

\bibitem[{{Lavalley} {et~al.}(1992){Lavalley}, {Isobe}, \&
  {Feigelson}}]{lif92a}
{Lavalley}, M.~P., {Isobe}, T., \& {Feigelson}, E.~D. 1992, in BAAS, Vol.~24,
  839--840

\bibitem[{{Lovell} {et~al.}(1996){Lovell}, {Reynolds}, {Jauncey}, {Backus},
  {McCulloch}, {Sinclair}, {Wilson}, {Tzioumis}, {King}, {Gough}, {Ellingsen},
  {Phillips}, {Preston}, \& {Jones}}]{lrj+96}
{Lovell}, J.~E.~J., {Reynolds}, J.~E., {Jauncey}, D.~L., {et~al.} 1996, ApJ,
  472, L5

\bibitem[{{Maccagni} {et~al.}(2017){Maccagni}, {Morganti}, {Oosterloo},
  {Ger{\'e}b}, \& {Maddox}}]{mmo+17}
{Maccagni}, F.~M., {Morganti}, R., {Oosterloo}, T.~A., {Ger{\'e}b}, K., \&
  {Maddox}, N. 2017, A\&A, in press (arXiv:1705.00492)

\bibitem[{{Madau} \& {Dickinson}(2014)}]{md14}
{Madau}, P. \& {Dickinson}, M. 2014, Ann. Rev. Astr. Ap., 52, 415

\bibitem[{{Micha{\l}owski} {et~al.}(2015){Micha{\l}owski}, {Gentile}, {Hjorth},
  {Krumholz}, {Tanvir}, {Kamphuis}, {Burlon}, {Baes}, {Basa}, {Berta}, {Castro
  Cer{\'o}n}, {Crosby}, {D'Elia}, {Elliott}, {Greiner}, {Hunt}, {Klose},
  {Koprowski}, {Le Floc'h}, {Malesani}, {Murphy}, {Nicuesa Guelbenzu},
  {Palazzi}, {Rasmussen}, {Rossi}, {Savaglio}, {Schady}, {Sollerman}, {de
  Ugarte Postigo}, {Watson}, {van der Werf}, {Vergani}, \& {Xu}}]{mgh+15}
{Micha{\l}owski}, M.~J., {Gentile}, G., {Hjorth}, J., {et~al.} 2015, A\&A, 582,
  A78

\bibitem[{{Morganti} {et~al.}(2015){Morganti}, {Sadler}, \& {Curran}}]{msc+15}
{Morganti}, R., {Sadler}, E.~M., \& {Curran}, S. 2015, Advancing Astrophysics
  with the Square Kilometre Array (AASKA14), 134

\bibitem[{{Neeleman} {et~al.}(2016){Neeleman}, {Prochaska}, {Ribaudo},
  {Lehner}, {Howk}, {Rafelski}, \& {Kanekar}}]{npr+16}
{Neeleman}, M., {Prochaska}, J.~X., {Ribaudo}, J., {et~al.} 2016, ApJ, 818, 113

\bibitem[{{Noterdaeme} {et~al.}(2012){Noterdaeme}, {Petitjean}, {Carithers},
  {P{\^a}ris}, {Font-Ribera}, {Bailey}, {Aubourg}, {Bizyaev}, {Ebelke},
  {Finley}, {Ge}, {Malanushenko}, {Malanushenko}, {Miralda-Escud{\'e}},
  {Myers}, {Oravetz}, {Pan}, {Pieri}, {Ross}, {Schneider}, {Simmons}, \&
  {York}}]{npc+12}
{Noterdaeme}, P., {Petitjean}, P., {Carithers}, W.~C., {et~al.} 2012, A\&A,
  547, L1

\bibitem[{{O'Dea} {et~al.}(1994){O'Dea}, {Baum}, \& {Gallimore}}]{obg94}
{O'Dea}, C.~P., {Baum}, S.~A., \& {Gallimore}, J.~F. 1994, ApJ, 436, 669

\bibitem[{Peacock(1999)}]{pea99}
Peacock, J.~A. 1999, Cosmological Physics (Cambridge: Cambridge University
  Press)

\bibitem[{Prochaska \& Herbert-Fort(2004)}]{ph04}
Prochaska, J.~X. \& Herbert-Fort, S. 2004, PASP, 116, 622

\bibitem[{Prochaska {et~al.}(2005)Prochaska, Herbert-Fort, \& Wolfe}]{phw05}
Prochaska, J.~X., Herbert-Fort, S., \& Wolfe, A.~M. 2005, ApJ, 635, 123

\bibitem[{{Prochaska} \& {Wolfe}(2009)}]{pw09}
{Prochaska}, J.~X. \& {Wolfe}, A.~M. 2009, ApJ, 696, 1543

\bibitem[{{Purcell} \& {Field}(1956)}]{pf56}
{Purcell}, E.~M. \& {Field}, G.~B. 1956, ApJ, 124, 542

\bibitem[{Rao {et~al.}(2006)Rao, Turnshek, \& Nestor}]{rtn05}
Rao, S., Turnshek, D., \& Nestor, D.~B. 2006, ApJ, 636, 610

\bibitem[{{Rao} \& {Turnshek}(2000)}]{rt00}
{Rao}, S.~M. \& {Turnshek}, D.~A. 2000, ApJS, 130, 1

\bibitem[{{Rao} {et~al.}(2017){Rao}, {Turnshek}, {Sardane}, \&
  {Monier}}]{rtsm17}
{Rao}, S.~M., {Turnshek}, D.~A., {Sardane}, G.~M., \& {Monier}, E.~M. 2017,
  MNRAS, submitted (arXiv:1704.01634)

\bibitem[{{Rhee} {et~al.}(2013){Rhee}, {Zwaan}, {Briggs}, {Chengalur}, {Lah},
  {Oosterloo}, \& {van der Hulst}}]{rzb+13}
{Rhee}, J., {Zwaan}, M.~A., {Briggs}, F.~H., {et~al.} 2013, MNRAS, 435, 2693

\bibitem[{{Roy} {et~al.}(2013){Roy}, {Mathur}, {Gajjar}, \& {Nath
  Patra}}]{rmgn13}
{Roy}, N., {Mathur}, S., {Gajjar}, V., \& {Nath Patra}, N. 2013, MNRAS, 436,
  L94

\bibitem[{{Sobral} {et~al.}(2013){Sobral}, {Smail}, {Best}, {Geach}, {Matsuda},
  {Stott}, {Cirasuolo}, \& {Kurk}}]{ssb+13}
{Sobral}, D., {Smail}, I., {Best}, P.~N., {et~al.} 2013, MNRAS, 428, 1128

\bibitem[{{Spring} \& {Micha{\l}owski}(2017)}]{sm17}
{Spring}, E.~F. \& {Micha{\l}owski}, M.~J. 2017, MNRAS, submitted
  (arXiv:1707.08877)

\bibitem[{{Srianand} {et~al.}(2012){Srianand}, {Gupta}, {Petitjean},
  {Noterdaeme}, {Ledoux}, {Salter}, \& {Saikia}}]{sgp+12}
{Srianand}, R., {Gupta}, N., {Petitjean}, P., {et~al.} 2012, MNRAS, 421, 651

\bibitem[{{Staveley-Smith} \& {Oosterloo}(2015)}]{so15}
{Staveley-Smith}, L. \& {Oosterloo}, T. 2015, Advancing Astrophysics with the
  Square Kilometre Array (AASKA14), 167

\bibitem[{{Webster} {et~al.}(1995){Webster}, {Francis}, {Peterson},
  {Drinkwater}, \& {Masci}}]{wfp+95}
{Webster}, R.~L., {Francis}, P.~J., {Peterson}, B.~A., {Drinkwater}, M.~J., \&
  {Masci}, F.~J. 1995, Nat, 375, 469

\bibitem[{{Wolfe} {et~al.}(2005){Wolfe}, {Gawiser}, \& {Prochaska}}]{wgp05}
{Wolfe}, A.~M., {Gawiser}, E., \& {Prochaska}, J.~X. 2005, ARA\&A, 43, 861

\bibitem[{{Wolfire} {et~al.}(1995){Wolfire}, {Hollenbach}, {McKee}, {Tielens},
  \& {Bakes}}]{whm+95}
{Wolfire}, M.~G., {Hollenbach}, D., {McKee}, C.~F., {Tielens}, A.~G.~G.~M., \&
  {Bakes}, E.~L.~O. 1995, ApJ, 443, 152

\bibitem[{York {et~al.}(2007)York, Kanekar, Ellison, \& Pettini}]{ykep07}
York, B.~A., Kanekar, N., Ellison, S.~L., \& Pettini, M. 2007, MNRAS, 382, L53

\bibitem[{{Zwaan} {et~al.}(2015){Zwaan}, {Liske}, {P{\'e}roux}, {Murphy},
  {Bouch{\'e}}, {Curran}, \& {Biggs}}]{zlp+15}
{Zwaan}, M.~A., {Liske}, J., {P{\'e}roux}, C., {et~al.} 2015, MNRAS, 453, 1268

\bibitem[{{Zwaan} {et~al.}(2005){Zwaan}, {van der Hulst}, {Briggs},
  {Verheijen}, \& {Ryan-Weber}}]{zvb+05}
{Zwaan}, M.~A., {van der Hulst}, J.~M., {Briggs}, F.~H., {Verheijen}, M.~A.~W.,
  \& {Ryan-Weber}, E.~V. 2005, MNRAS, 364, 1467

\bibitem[{{Zwart} {et~al.}(2014){Zwart}, {Jarvis}, {Deane}, {Bonfield},
  {Knowles}, {Madhanpall}, {Rahmani}, \& {Smith}}]{zjd+14}
{Zwart}, J.~T.~L., {Jarvis}, M.~J., {Deane}, R.~P., {et~al.} 2014, MNRAS, 439,
  1459

\end{thebibliography}

\end{document}